\documentclass[preprint,amsmath,amssymb]{revtex4}
\usepackage{graphicx}
\usepackage{dcolumn}
\usepackage{bm}
\usepackage{natbib}

\begin{document}
\title{Coarse-grained strain dynamics and backwards/forwards dispersion}
\author{Jacob Berg}
\email{jacob.berg.joergensen@risoe.dk}
\author{Beat L\"{u}thi}
\author{S{\o}ren Ott}
\author{Jakob Mann}

\affiliation{Ris{\o} National Laboratory, 4000 Roskilde, Denmark}

\date{\today}
\begin{abstract}
A Particle Tracking Velocimetry experiment has been performed in a
turbulent flow at intermediate Reynolds number. We present
experimentally obtained stretching rates for particle pairs in the
inertial range. When compensated by a characteristic time scale
for coarse-grained strain we observe constant stretching. This
indicates that the process of material line stretching taking
place in the viscous subrange has its counterpart in the inertial
subrange. We investigate both forwards and backwards dispersion.
We find a faster backwards stretching and relate it to the problem
of relative dispersion and its time asymmetry.
\end{abstract}

\maketitle

\section{Introduction}
The time asymmetry existing in a turbulent flow governed by the
Navier-Stokes equation has recently been discussed in the
connection with the diffusion properties of two nearby fluid
particles initially close to each other. A stochastic model study
by \citet{sawford:2005} showed that backwards dispersion is faster
than the corresponding forwards dispersion. This has practical
implications for a variety of applications dealing with transport
and mixing in turbulent flows.

The asymmetry was experimentally verified by a Particle Tracking
Velocimetry (PTV) experiment by \citet{berg:2006} who speculated
that kinematic infinitesimal material line stretching has its
counter-part in the inertial range. Different stretching rates
could then explain the observed difference in forwards and
backwards dispersion. Through a finite Reynolds number formulation
of the much celebrated Richardson law $\langle r^2 \rangle =
g\varepsilon t^3$ it was found that following two particles
backwards in time the dispersion was faster by a factor of
approximately two compared with following the particles forwards
in time. This ratio was also obtained from Direct Numerical
Simulation (DNS) analysis by the same authors (data originally
presented in \citep{biferale:2004,biferale:2005b}).
\citet{sawford:2005} found the same ratio to be much larger:
between $5$ and $20$.

Yet another stochastic model was developed by \citet{luthi:2006b}.
By assuming that stretching rates behave self-similar results
consistent with \cite{berg:2006} were obtained. By self-similar
stretching we think of particle separation which is similar on the
different scales. The self-similarity of stretching rates has,
however, not been shown.

In this paper we will explore forwards/backwards dispersion in the
context of coarse-grained strain dynamics. The latter has recently
got a lot of attention in the turbulence community since it is
more or less related to Large Eddy Simulations (LES)
\citep{borue:1998,chertkov:1999,tao:2002,bos:2002,pumir:2003,naso:2006,luthi:2006b}.
One can expect Kolmogorov similarity scaling to hold for the
coarse-grained quantities as long as the filtering scale is less
than the integral scale in the flow where the turbulence is solely
represented by the kinetic energy dissipation of the flow,
$\varepsilon$

As in \cite{luthi:2006b} we define the coarse-grained velocity
gradient tensor to be
\begin{equation}
\widetilde{A}^{\Delta}_{ij}=\frac{\partial
\widetilde{u}^{\Delta}_j}{\partial x_i}, \label{eqn:cgA}
\end{equation}
where $\tilde{\mathbf{u}}^{\Delta}(\mathbf{x},t)$ is the
coarse-grained velocity over scale $\Delta$:
\begin{equation}
\widetilde{u}^{\Delta}_i(\mathbf{x})=\frac{1}{V}
\int_{B_{\Delta/2}} d\mathbf{r}\hspace{0.05cm}
u_i(\mathbf{x}+\mathbf{r}). \label{eqn:u}
\end{equation}
Here $B_{\Delta/2}$ denotes a Ball with radius $\Delta/2$ and $V$
its volume.

We will follow the lines of \cite{berg:2006} and
\citet{luthi:2006b} and present experimental evidence of
self-similar stretching in a turbulent flow of intermediate
Reynolds number. The result is linked to particle pair separation
and is able to account for the difference in forwards and
backwards dispersion.

\section{Experimental method}
\begin{figure}[h]
\includegraphics[width=0.8\columnwidth]{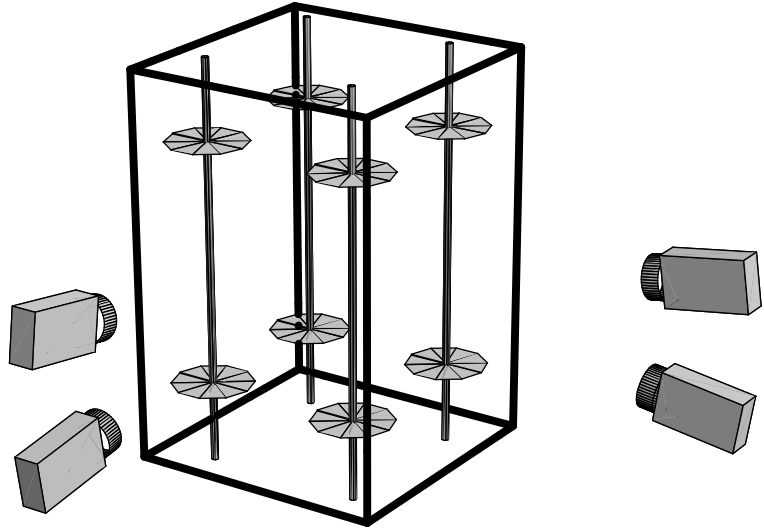}\vspace{-2cm}
\caption{{\footnotesize Experimental setup}} \label{fig:exp}
\end{figure}

We have performed a Particle Tracking Velocimetry (PTV) experiment
in an intermediate Reynolds number turbulent flow. The flow has
earlier been reported in \cite{berg:2006}. PTV is an experimental
method suitable for obtaining Lagrangian statistics in turbulent
flows
\citep{ott:2000,laporta:2001,mordant:2001,luthi:2005,cornell,xu:2006}:
Lagrangian trajectories of fluid particles in water are obtained
by tracking neutrally buoyant particles in space and time. The
flow is generated by eight rotating propellers, which change their
rotational direction in fixed intervals in order to suppress a
mean flow, placed in the corners of a tank with dimensions
$32\times32\times50\mathrm{cm}^3$ (see Fig~\ref{fig:exp}).
The data acquisition system consists of four commercial CCD
cameras with a maximum frame rate of $50\mathrm{Hz}$ at $1000
\times 1000$ pixels. The measuring volume covers roughly
$(12\mathrm{cm})^3$. We use polystyrene particles with size
$\sim400\mathrm{\mu m}$ and density very close to one. We follow
$\mathcal{O}(1000)$ particles at each time step with a position
accuracy of $0.05$ pixels corresponding to less than $10\mathrm{
\mu m}.$

The Stokes number, $\tau_I / \tau_{\eta}$ ($\tau_I$ denotes the
inertial relaxation time for the particle to the flow while
$\tau_{\eta}$ is the Kolmogorov time) is much less than one and
the particles can therefore be treated as passive tracers in the
flow. The particles are illuminated by a $250\mathrm{W}$ flash
lamp.

The mathematical algorithms for translating two dimensional image
coordinates from the four camera chips into a full set of three
dimensional trajectories in time involve several crucial steps:
fitting gaussian profiles to the 2d images, stereo matching (line
of sight crossings) with a two media (water-air) optical model and
construction of 3d trajectories in time by using the kinematic
principle of minimum change in acceleration
\citep{willneff:2003,ouellette:2006}.

\begin{table}[t]
\begin{tabular}{|c|c|c|c|c|c|c|}
  \hline
  $\eta$ & $L_{int}$ & $\tau_{\eta}$ & $T_L$ &$\varepsilon$ & $\sigma_u$ & $Re_{\lambda}$ \\
  \hline\hline
  $0.25 \mathrm{mm}$ & $48\mathrm{mm}$ & $0.07\mathrm{s}$ & $2.45\mathrm{s}$& $168\mathrm{mm^2/s^3}$ & $23.33\mathrm{mm/s}$ & 172 \\
  \hline
\end{tabular}
\caption{{\footnotesize Turbulence characteristics obtained from
the von K\'{a}rm\'{a}n model~\cite{ott:2000}: Mean kinetic energy
dissipation $\varepsilon$, Kolmogorov length
$\eta\equiv(\nu^3/\varepsilon)^{1/4}$, Kolmogorov time
$\tau_{\eta}\equiv (\nu/\varepsilon)^{1/2}$, integral length
$L_{int}$, integral time $T_L$, velocity fluctuations
$\sigma_u^2=\frac{1}{3}(\sigma_{u_x}^2+\sigma_{u_y}^2+\sigma_{u_z}^2).$
The Reynolds number is defined as $Re_{\lambda}=\frac{\lambda
\sigma_u}{\nu}$ with the Taylor micro scale
$\lambda=\sqrt{\left(\frac{15 \nu
\sigma^2}{\varepsilon}\right)}.$}} \label{table}
\end{table}

From eqn.~\ref{eqn:cgA} we can define the coarse grained strain
$\langle \widetilde{s}^2\rangle$ and
$\langle\widetilde{\Omega}^2\rangle$ where $\langle
\widetilde{s}_{ij} \rangle$ and $\langle
\widetilde{\Omega}_{ij}\rangle$ are the symmetric and
anti-symmetric part of the coarse-grained velocity gradient tensor
$\langle\widetilde{A}_{ij}\rangle$. The eigenvalues and
eigenvector of $\langle\widetilde{s}_{ij}\rangle$ is denoted
$\langle\widetilde{\Lambda}_i\rangle$ and $\langle
\widetilde{\lambda}_i \rangle$ respectively. Due to
incompressibility $\sum_i \langle \widetilde{\Lambda}_i
\rangle=0$. In addition we define
$\langle\widetilde{\Lambda}_1\rangle>\langle\widetilde{\Lambda}_2\rangle>\langle\widetilde{\Lambda}_3\rangle$
so that the most positive principal direction of strain is
$\langle\widetilde{\Lambda}_1\rangle$.

The filtered velocity field (eqn.~\ref{eqn:u}) is approximated by
least square fits of spherical incompressible and orthogonal
linear polynomials to discrete velocities of fluid particles
inside spherical balls with diameter $\Delta$:
\begin{equation}
\min \left[\int_{B(\Delta/2)} d^3\mathbf{x}
\hspace{0.1cm}(u_i-\widetilde{A}_{ij} x_j)^2\right]
\end{equation}
At least four particles are necessary to describe a
three-dimensional shape and hence to estimate $\langle
\widetilde{A}_{ij} \rangle.$ Four particles form a tetrahedron and
is the backbone of the analysis by \citet{chertkov:1999}. In
\cite{luthi:2006b} we show that $\widetilde{A}_{ij}$ obtained from
only four particles is quite far from the definition in
eqn.~\ref{eqn:u} and therefore that using only four particles are
not sufficient.

\begin{figure}[h]
\begin{center}
\includegraphics[width=\columnwidth]{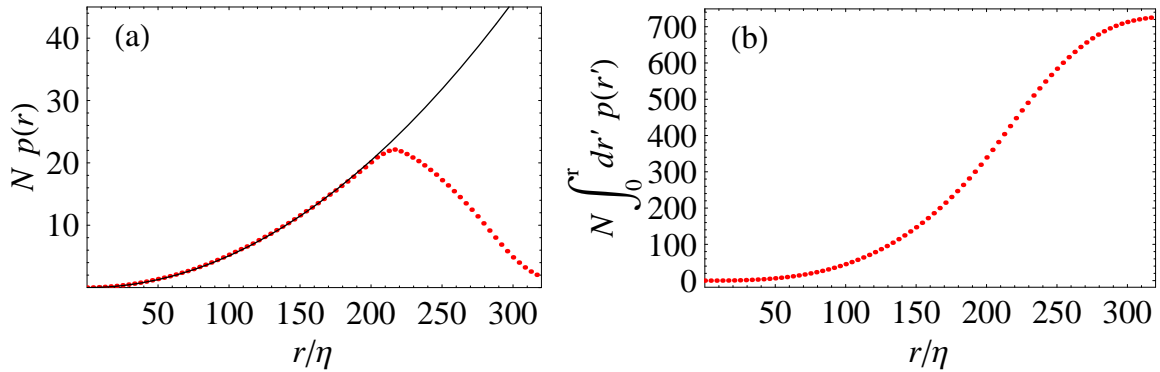}
\caption{{\footnotesize Radial distribution of particles where $N$
is the total number of particles in the volume. a) The probability
of finding a particle in a spherical shell with thickness $dr$ and
with distance $r$ from the center of the measuring volume. b)
Cumulative distribution of (a). }} \label{fig:distributions}
\end{center}
\end{figure}

We find that at least twelve particles are needed in order to
obtain a reasonable approximation of eqn.~\ref{eqn:u} of the
coarse-grained quantities. Using this many particles has the
drawback that we can not study the dynamics at the smallest
scales. In Figure \ref{fig:distributions} (a) the radial
distribution of particles is shown. The probability $N p(r)$ of
finding a given number $N$ of particles on the surface of a ball
centered in our measuring volume with radius $r$ is observed to
increase with $r^2$ up to $\sim200\eta$. This means that the
particle density can be considered uniform up to this scale. For
lager radius the density drops down because of non-perfect
illumination at the boundary of the measuring volume. The
cumulative distribution is show in (b) and is interesting because
it gives us an estimate of the number of particles we can expect
to find in a ball with radius $r$. The number is, however, only an
upper bound since the ball is centered in the measuring volume and
not around all individual particles in the flow.

The smallest scale for which $\langle \widetilde{u}\rangle$ and
hence $\langle\widetilde{s}^2\rangle$ and
$\langle\widetilde{\Omega}^2\rangle$ can be resolved is $80 \eta.$
A higher particle seeding density which again makes particle
tracking more difficult can decrease this number. All results
reported use a minimum of twelve particles.

\begin{figure}[h]
\begin{center}
\includegraphics[width=\columnwidth]{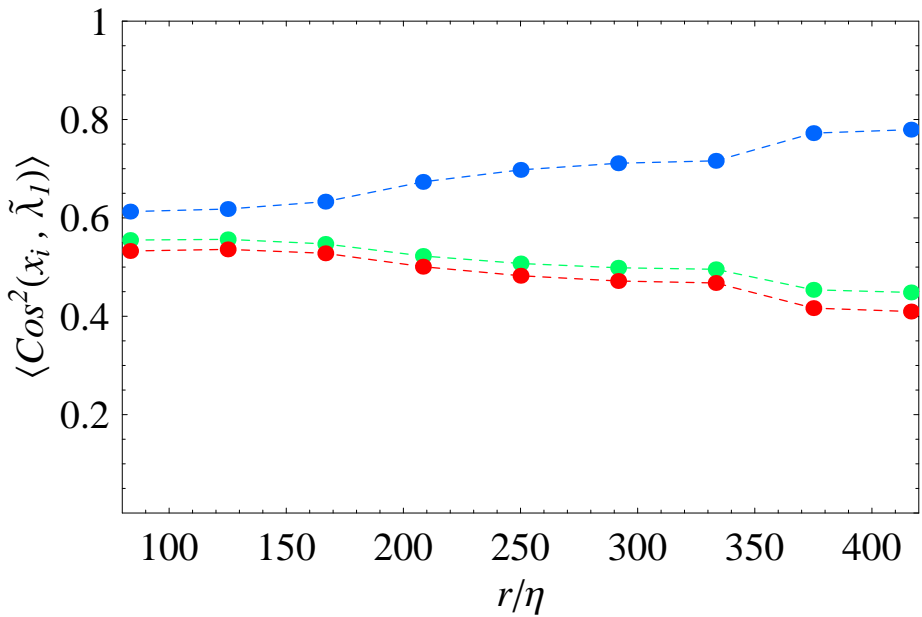}
\caption{{\footnotesize Squared cosine between the most stretching
principle coarse grained eigenvector, $\widetilde{\lambda}_1$, and
the three coordinate directions. Blue is the vertical
axis-symmetric direction while the green and red is the horizontal
directions. }} \label{fig:largeScaleStrain}
\end{center}
\end{figure}

As already reported in \cite{berg:2006,luthi:2006b} the mean flow
is slightly straining though with a characteristic time scale many
times larger than the integral time scale of the flow. An
alternative way of emphasizing  the large scale mean strain is
shown in Figure~\ref{fig:largeScaleStrain}.
$\cos^2(x_i,\widetilde{\lambda}_1),$ where $x_i$ is the three
coordinate axes, is observed to peak for large $r$ in the vertical
direction while the horizontal axes decrease in agreement with an
axis-symmetric flow. A slow convergence towards isotropy is
observed as $r=\Delta$ is reduced.

\section{Coarse-grained statistics}
In the viscous subrange the stretching rate of infinitesimal
material lines governs particle pair separation. The stretching
rate is defined as
\begin{equation}
L(t)=\left \langle \frac{1}{2 r^2} \frac{d r^2}{dt}\right \rangle
\label{eq:L}.
\end{equation}
where $d/dt$ denotes Lagrangian differentiation (following the
particles). In the viscous subrange $L(t)$ when rescaled with the
Kolmogorov time scale $\tau_{\eta}$ becomes constant after a short
time~\citep{biferale:2005,girimaji:1990,guala:2005,berg:2006,luthi:2005}.
In the viscous sub range the second order Eulerian longitudinal
structure function $f(r)=\langle
\{[u_i(\mathbf{x}+\mathbf{r})-u_i(\mathbf{x})] r_i/r\}^2 \rangle$
is given by
\begin{equation}
f(r)=\frac{\langle s^2\rangle}{15 }r^2 \hspace{1cm} r\ll
\eta\label{eq:f}
\end{equation}
Through the definitions $\varepsilon=2 \nu \langle s^2 \rangle$
and $\tau_{\eta}^2=\nu/\varepsilon$ we can form
$\tau_{\eta}^2=r^2/(30 f(r)).$ Thus motivated by viscous subrange
scaling we move to the inertial range and define a time scale
\begin{equation}
t_{\star}(r)=\sqrt{\frac{2 r^2}{15 f(r)}} \hspace{1cm} r \ll
L_{int}. \label{eq:tstar}
\end{equation}
In the limit $r\ll \eta$ we have $t_{\star}=\sqrt{2} \tau_{\eta}$.

\begin{figure}[h]
\begin{center}
\includegraphics[width=\columnwidth]{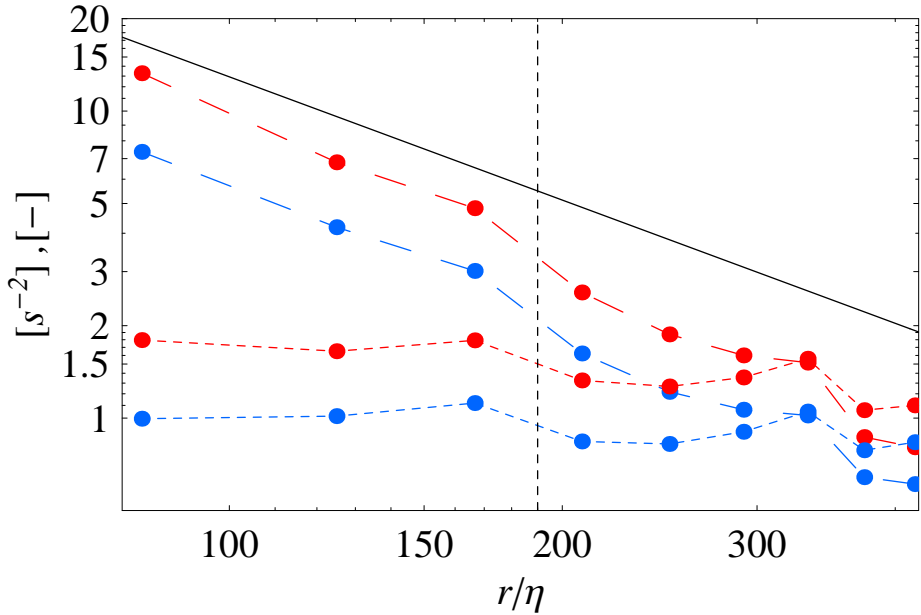}
\caption{{\footnotesize Coarse grained strain $\langle
\widetilde{s}^2 \rangle$ (long dashed blue) and $\langle
\widetilde{\Omega}^2 \rangle$ (long dashed red) as a function of
coarse graining scale $r.$ The solid black line is $r^{-4/3}$. The
dotted lines are $\langle \widetilde{s}^2 \rangle$ and $\langle
\widetilde{\Omega}^2 \rangle$ multiplied by $t_{\star}^2.$ The
vertical black dotted line indicates the integral length scale.}}
\label{fig:strainAndEnstrophy}
\end{center}
\end{figure}

As a function of scale $r=\Delta$ we plot $\langle \widetilde{s}^2
\rangle$ and $\langle \tilde{\Omega}^2\rangle $ in
Figure~\ref{fig:strainAndEnstrophy}. Both quantities are in the
inertial range observed to be in agreement with the Kolmogorov
similarity prediction $r^{-4/3}.$ For $r \rightarrow 0$ the ratio
$\langle \widetilde{\Omega}^2 \rangle /\langle \tilde{s}^2\rangle
\rightarrow 2$.

Also in Figure~\ref{fig:strainAndEnstrophy} we have plotted
$\langle \widetilde{s}^2 \rangle$ and $\langle
\widetilde{\Omega}^2 \rangle$ multiplied with $t^2_{\star}$. In
the inertial range we find that $\langle \widetilde{s}^2 \rangle
t^2_\star \sim 1$ and $\langle \widetilde{\Omega}^2 \rangle
t^2_\star = 1.85.$  It thus seems that $t_{\star}$ as defined in
eqn.~\ref{eq:tstar} serves as a characteristic time scale for
$\langle \widetilde{s}^2\rangle$ and therefore
$t_{\star}\sim1/\sqrt{\langle \widetilde{s}^2 \rangle}$.

In \cite{luthi:2006b} $f(r)$ in eqn.~\ref{eq:tstar} was defined
through a model valid for homogeneous and isotropic turbulence. In
this paper $f(r)$ are obtained directly from data. The behavior
therefore is therefore slightly different.

\section{Particle pair separation}
\begin{figure}[h]
\begin{center}
\includegraphics[width=\columnwidth]{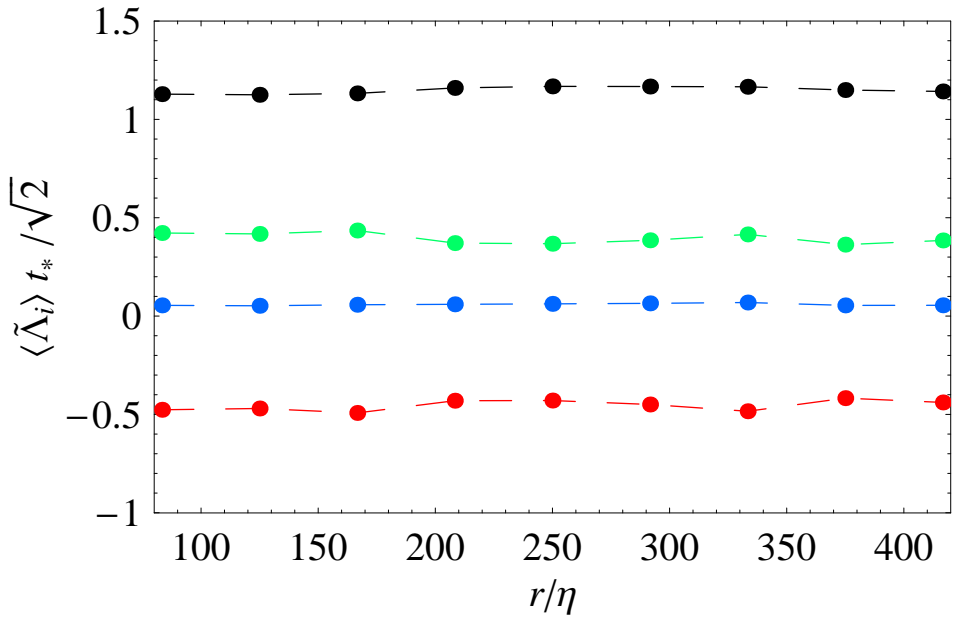}
\caption{{\footnotesize Eigenvalues $\langle \widetilde{\Lambda}_i
\rangle t_{\star}$ of $\langle \widetilde{s}_{ij}\rangle$. $i=1$:
green, $i=2$: blue and $i=3$: red. The black curve is the ratio $|
\langle \widetilde{\Lambda}_3 \rangle|/|\langle
\widetilde{\Lambda}_1\rangle|$. }} \label{fig:Lambda}
\end{center}
\end{figure}

The rescaled eigenvalues $\langle \widetilde{\Lambda}_i \rangle
t_{\star}$ are shown in Figure~\ref{fig:Lambda}. The trademark of
turbulence, namely $\langle \widetilde{\Lambda}_2 \rangle>0$ which
is necessary for both positive mean enstrophy and strain
production is observed for all values of $r$. A slight decrease in
$\langle \widetilde{\Lambda}_2 \rangle$ is observed as $r$ is
increased which could be taken as a sign of the coarse-graining
field becoming more Gaussian and hence
$\langle\widetilde{\Lambda}_2\rangle=0.$ Both
$\langle\widetilde{\Lambda}_1\rangle t_{\star}$ and
$\langle\widetilde{\Lambda}_3\rangle t_{\star}$ are constant and
so are their ratio
$|\langle\widetilde{\Lambda}_3\rangle|/|\langle\widetilde{\Lambda}_1\rangle|$.
For a direct comparison with viscous result we have divided the
results with $\sqrt{2}$.

\begin{figure}[h]
\begin{center}
\includegraphics[width=\columnwidth]{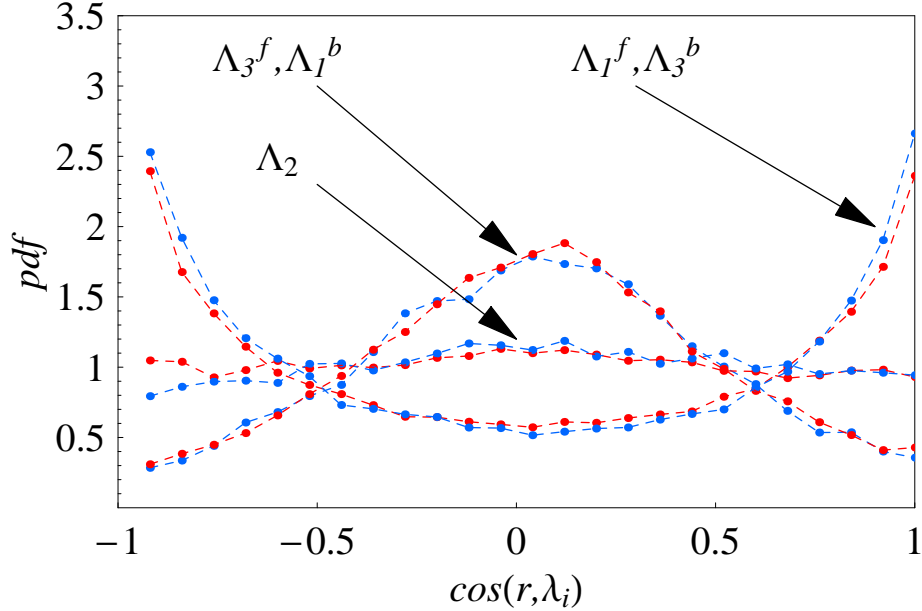}
\caption{{\footnotesize Angular dependence between $\langle
\widetilde{\lambda_i} \rangle$ and the separation vector $r$ for
times $t_B\gg t \gg 5t_B$. Red curves are forwards dispersion
while blue are backwards dispersion.}} \label{fig:cosRL}
\end{center}
\end{figure}

When time is running forward two particles in a mean strain field
will on average separate from each other along the direction of
$\langle \widetilde{\lambda}_1\rangle$. In the backward case they
will separate along the direction of
$\langle\widetilde{\lambda}_3\rangle$. Since
$\langle\widetilde{\Lambda}_2\rangle>0$ and therefore
$|\langle\widetilde{\Lambda}_3 \rangle|>|\langle
\widetilde{\Lambda}_1\rangle|$ backwards separation is the faster
one. This is shown in Figure~\ref{fig:cosRL} for times $t_B<t<
5t_B$ where $t_B=(r^2_0 / \varepsilon)^{1/3}$ is the Batchelor
time, characterizing the time for which the initial separation
should be regarded an important parameter in the separation
process~\cite{cornell}.
By closer inspection of Figure~\ref{fig:cosRL} we can see that in
the backward case $r$ is slightly more aligned with
$\langle\widetilde{\lambda}_3\rangle$ than it is aligned with
$\langle\widetilde{\lambda}_1\rangle$ in the forward case. This is
due to the positiveness of $\langle\widetilde{\Lambda}_2\rangle$
and a corresponding increase in the alignment of $r$ with
$\langle\widetilde{\lambda}_2\rangle$ in the forward case compared
to in the backward case is therefore also observed. The values of
$\cos^2{(r,\widetilde{\lambda}_i)}$ are in the forward case
$0.50$, $0.32$ and $0.19$ for $i=1,2,3$ respectively. In the
backward case the same values are $0.19$, $0.28$ and $0.53$.

\begin{figure}[h]
\begin{center}
\includegraphics[width=\columnwidth]{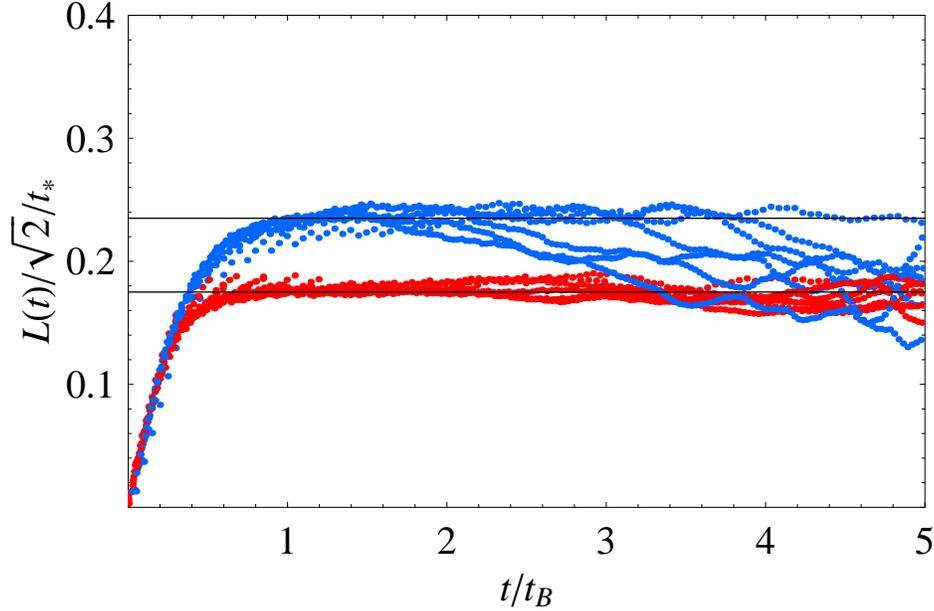}
\caption{{\footnotesize Stretching rates $L(t)$ rescaled by
$t_{\star}$. The blue curves correspond to forwards separation
while the red curves corresponds to backwards separation (time
running backwards) for different initial separations $r_0$.
Particle pairs with $r_0$ from $3-4\mathrm{mm}$ to
$9-10\mathrm{mm}$ in steps of $1mm$ are shown. We have divided the
$y$-axis with $\sqrt{2}$ in order to match the viscous limit.}}
\label{fig:sr}
\end{center}
\end{figure}

Stretching rates rescaled by $t_{\star}$ are presented in
Figure~\ref{fig:sr} as a function of $t/t_B$ for different initial
separation of pairs.

For times $t\sim 0.5 t_B$ the forward stretching and backward
stretching occurs with the same speed. The separation vector is
still randomly oriented and has therefore not yet aligned itself
with the principal directions of the strain field. For longer
times the backwards stretching rates saturates at $0.235\pm0.005$
while the forwards saturate at $0.18\pm0.005$. Whereas all curves
in the forward case collapse for times up to $3-4 t_B$ the curves
in the backward case do not collapse so nicely. Because backward
separation is faster the particles leave the measurement volume
earlier: the larger $r_0$, the earlier. The particle pairs we thus
observe for large times are likely to be slow pairs with low
stretching rate. To reduce the effect of finite volume we only
choose pairs which start within a small sub volume
($30\mathrm{mm}$) of the full measuring volume. There is, however,
no systematic way in which we can totally neglect the effect of a
finite volume.

The maximum stretching rates would occur if the particle
separation vector $\mathbf{r}$ is fully aligned with the principal
strain axes. In this case the rescaled stretching rates would be
$\langle \widetilde{\Lambda}_1 \rangle t_{\star}$ and $-\langle
\widetilde{\Lambda}_3 \rangle t_{\star}$ in
Figure~\ref{fig:Lambda} for the forward and backward case
respectively. The forward value
$L_f(t)/\sqrt{2}/t_{\star}=0.18\pm0.005$ is close to the values
obtained in the viscous sub
range~\cite{biferale:2005,girimaji:1990,guala:2005,berg:2006}.

\citet{berg:2006} hypothesized that, if the particle separation
was perfectly aligned with the principal strain axes, the ratio
between forwards and backwards dispersion rates characterized by
the Richardson-Obukhov constant ratio $g_b/g_f$ could be
determined as $|\langle \widetilde{\Lambda}_3 \rangle|/\langle
\widetilde{\Lambda}_1 \rangle)^3$. From Figure~\ref{fig:Lambda}
this number equals $1.13^3=1.44$. From the calculated stretching
rates we find that this is certainly not the case since
$0.235/0.18=1.31$. With the same argumentation this would give a
ratio $g_b/g_f=2.23\pm0.33$ within errors of what was measured
directly in \cite{berg:2006}. Although
$L(t)_f\le|\langle\widetilde{\Lambda}_1 \rangle|$ and
$L(t)_b\le|\langle\widetilde{\Lambda}_3\rangle|$ we thus have
$L(t)_b/L(t)_f\le|\langle\widetilde{\Lambda}_3 \rangle/\langle
\widetilde{\Lambda}_1\rangle|$ and hence a larger ratio than in
the fully aligned case. It is important to stress that the
argumentation of how to relate the ratio $g_b/g_f$ to
$L(t)_b/L(t)_f$ through the Richardson-Obukhov law is not a
rigourously derived result but merely a hand waving argument.

\section{Conclusions}
We have given evidence of the existence of, what we call
self-similar stretching rates: when scaled by a relevant time
$t_{\star}$ scale which is a function of the second order
structure function stretching rates of infinitesimal material
lines has its counterpart in the inertial range. Furthermore it
turns out that this relevant time scale $t_{\star}$ is related to
the coarse-grained strain field. The stretching is like in the
viscous range far from being perfectly aligned with the most
positive direction of strain which we have shown would lower the
ratio between forwards and backwards dispersion.

Whether or not the Lagrangian stretching rates found in this paper
are universal or specific for this particular turbulent flow other
experiments and/or DNS will have to decide.

\bibliographystyle{plainnat}

\begin{thebibliography}{22}
\expandafter\ifx\csname
natexlab\endcsname\relax\def\natexlab#1{#1}\fi
\expandafter\ifx\csname url\endcsname\relax
  \def\url#1{{\tt #1}}\fi

\bibitem[Berg et~al.(2006)Berg, {L\"{u}thi}, Mann, and Ott]{berg:2006}
J.~Berg, B.~{L\"{u}thi}, J.~Mann, and S.~Ott.
\newblock Backwards and forwards relative dispersion in turbulent flow: An
  experimental investigation.
\newblock {\em Phys.\ Rev.\ E}, 34:\penalty0 115, 2006.

\bibitem[Biferale et~al.(2004)Biferale, Boffetta, Celani, Devenish, Lanotte,
  and Toschi]{biferale:2004}
L.~Biferale, G.~Boffetta, A.~Celani, B.~J. Devenish, A.~Lanotte,
and F.~Toschi.
\newblock Multifractal statistics of lagrangian velocity and acceleration in
  turbulence.
\newblock {\em Phys.\ Rev.\ Lett.}, 93:\penalty0 064502, 2004.

\bibitem[Biferale et~al.(2005{\natexlab{a}})Biferale, Boffetta, Celani,
  Devenish, Lanotte, and Toschi]{biferale:2005}
L.~Biferale, G.~Boffetta, A.~Celani, B.~J. Devenish, A.~Lanotte,
and F.~Toschi.
\newblock Lagrangian statistics of particle pairs in homogeneous isotropic
  turbulence.
\newblock {\em Phys.\ Fluids}, 17:\penalty0 115101, 2005{\natexlab{a}}.

\bibitem[Biferale et~al.(2005{\natexlab{b}})Biferale, Boffetta, Celani,
  Devenish, Lanotte, and Toschi]{biferale:2005b}
L.~Biferale, G.~Boffetta, A.~Celani, B.~J. Devenish, A.~Lanotte,
and F.~Toschi.
\newblock Particle trapping in three-dimensional fully developed turbulence.
\newblock {\em Phys.\ Fluids}, 17:\penalty0 021701, 2005{\natexlab{b}}.

\bibitem[Borue and Orszag(2002)]{borue:1998}
V.~Borue and S.~A. Orszag.
\newblock Local energy flux and subgrid scale statistics in three-dimensional
  turbulence.
\newblock {\em J.\ Fluid \ Mech.}, 366:\penalty0 1, 2002.

\bibitem[Bourgoin et~al.(2006)Bourgoin, Ouellette, Xu, Berg, and
  Bodenschatz]{cornell}
M.~Bourgoin, N.~T. Ouellette, H.~Xu, J.~Berg, and E.~Bodenschatz.
\newblock The role of pair dispersion in turbulent flow.
\newblock {\em Science}, 311:\penalty0 835, 2006.

\bibitem[Chertkov et~al.(1999)Chertkov, Pumir, and Shraiman]{chertkov:1999}
M.~Chertkov, A.~Pumir, and B.~I. Shraiman.
\newblock Lagrangian tetrad dynamics and the phenomenology of turbulence.
\newblock {\em Phys.\ Fluids}, 11:\penalty0 2394, 1999.

\bibitem[Girimaji and Pope(1990)]{girimaji:1990}
S.~S. Girimaji and S.~B. Pope.
\newblock Material-element deformation in isotropic turbulence.
\newblock {\em J.\ Fluid Mech.}, 220:\penalty0 427, 1990.

\bibitem[Guala et~al.(2005)Guala, {L\"{u}thi}, Liberzon, Tsinober, and
  Kinzelbach]{guala:2005}
M.~Guala, B.~{L\"{u}thi}, A.~Liberzon, A.~Tsinober, and
W.~Kinzelbach.
\newblock On the evolution of material lines and vorticity in homogeneous
  turbulence.
\newblock {\em J.\ Fluid Mech.}, 533:\penalty0 339, 2005.

\bibitem[{L\"{u}thi} et~al.(2006){L\"{u}thi}, Berg, Ott, and J]{luthi:2006b}
B.~{L\"{u}thi}, J.~Berg, S.~Ott, and Mann J.
\newblock Lagrangian multi-particle statistics.
\newblock {\em \textit{submitted}}, 2006.

\bibitem[{L\"{u}thi} et~al.(2005){L\"{u}thi}, Tsinober, and
  Kinzelbach]{luthi:2005}
B.~{L\"{u}thi}, A.~Tsinober, and W.~Kinzelbach.
\newblock Lagrangian measurements of vorticity dynamics in tubulent flow.
\newblock {\em J.\ Fluid Mech.}, 528:\penalty0 87, 2005.

\bibitem[Mordant et~al.(2001)Mordant, Metz, Michel, and Pinton]{mordant:2001}
N.~Mordant, P.~Metz, O.~Michel, and J.-F. Pinton.
\newblock Mearesurement of lagrangian velocity in fully developed turbulence.
\newblock {\em Phys.\ Rev.\ Lett.}, 87:\penalty0 214501, 2001.

\bibitem[Naso et~al.(2006)Naso, Chertkov, and Pumir]{naso:2006}
A.~Naso, M.~Chertkov, and A.~Pumir.
\newblock Scale dependence of the coarse-grained velocity derivative tensor:
  influence of large-scale shear on small-scale turbulence.
\newblock {\em J.\ turbulence}, 7:\penalty0 41, 2006.

\bibitem[Ott and Mann(2000)]{ott:2000}
S.~Ott and J.~Mann.
\newblock An experimental investigation of the relative diffusion of particle
  pairs in three-dimensional flow.
\newblock {\em J.\ Fluid Mech.}, 422:\penalty0 207, 2000.

\bibitem[Ouellette et~al.(2006)Ouellette, Xu, and Bodenschatz]{ouellette:2006}
N.~T. Ouellette, H.~Xu, and E.~Bodenschatz.
\newblock A quantitative study of three-dimensional lagrangian particle
  tracking algorithms.
\newblock {\em Exp.\ in Fluids.}, 40:\penalty0 301, 2006.

\bibitem[Porta et~al.(2001)Porta, Voth, A.~M.~Crawford, and
  Bodenschatz]{laporta:2001}
A.~La Porta, G.~A. Voth, J.~Alexander A.~M.~Crawford, and
E.~Bodenschatz.
\newblock Fluid particle accelerations in fully developed turbulence.
\newblock {\em Nature}, 409:\penalty0 1017, 2001.

\bibitem[Pumir and Shraiman(2003)]{pumir:2003}
A.~Pumir and B.~I. Shraiman.
\newblock Lagrangian particle approach to large eddy simulations of
  hydrodynamic turbulence.
\newblock {\em J.\ Stat.\ Phys.}, 113:\penalty0 693, 2003.

\bibitem[Sawford et~al.(2005)Sawford, Yeung, and Borgas]{sawford:2005}
B.~L. Sawford, P.~K. Yeung, and M.~S. Borgas.
\newblock Comparison of backwards and forwards relative dispersion in
  turbulence.
\newblock {\em Phys.\ Fluids}, 17:\penalty0 095109, 2005.

\bibitem[Tao et~al.(2002)Tao, Katz, and Meneveau]{tao:2002}
B.~Tao, J.~Katz, and C.~Meneveau.
\newblock Statistical geometry of subgrid-scale stresses determined from
  holographic particle image velocimetry measurements.
\newblock {\em J.\ Fluid Mech.}, 457:\penalty0 35, 2002.

\bibitem[van~der Bos et~al.(2002)van~der Bos, Tao, and
  C.~Meneveau~and]{bos:2002}
F.~van~der Bos, B.~Tao, and J.~Katz C.~Meneveau~and.
\newblock Effects of small-scale turbulent motions on the altered velocity
  gradient tensor as deduced from holographic particle image velocimetry
  measurements.
\newblock {\em Phys.\ Fluids}, 14:\penalty0 2456, 2002.

\bibitem[Willneff(2003)]{willneff:2003}
J.~Willneff.
\newblock {\em A spatio-temporal mathing algorithm for 3D-particle tracking
  velocimetry}.
\newblock PhD thesis, ETH {Z\"{u}rich}, 2003.

\bibitem[Xu et~al.(2006)Xu, Bourgoin, Ouellette, and Bodenschatz]{xu:2006}
H.~Xu, M.~Bourgoin, N.~T. Ouellette, and E.~Bodenschatz.
\newblock High order lagrangian velocity statistics in turbulence.
\newblock {\em Phys.\ Rev.\ Lett.}, 96:\penalty0 024503, 2006.

\end{thebibliography}

\end{document}